\documentclass[a4paper,11pt,titlepage]{article}
\pdfoutput=1
\usepackage{natbib}
\usepackage{fullpage}       
\usepackage{graphicx}
\usepackage{times}
\usepackage{setspace}
\usepackage{tikz}
\usepackage{longtable}
\usepackage{rotating}
\usepackage{footnote}
\usepackage{amsmath}    
\usepackage{lscape}
\usepackage{afterpage}
\usepackage{placeins}
\usepackage{color}
\usepackage{url}
\usepackage{xr}
\usetikzlibrary{shapes,arrows}
\makesavenoteenv{tabular}
\begin{document}

\title{VSEAMS: A pipeline for variant set enrichment analysis using summary GWAS data identifies \textit{IKZF3}, \textit{BATF} and \textit{ESRRA} as key transcription factors in type 1 diabetes}
\author{Oliver S Burren\thanks{JDRF/Wellcome Trust Diabetes and Inflammation Laboratory, Cambridge Institute for Medical Research, University of Cambridge
WT/MRC Building, Cambridge, CB2 0XY, UK}, Hui Guo\footnotemark[1] and Chris Wallace\,\footnotemark[1]\,\thanks{To whom correspondence should be addressed (chris.wallace@cimr.cam.ac.uk)}}
\date{}

\maketitle

\section*{Abstract}

\textbf{Motivation}: Genome-wide association studies (GWAS) have identified many loci implicated in disease susceptibility.  Integration of GWAS summary statistics (p values) and functional genomic datasets should help to elucidate mechanisms.\\
\textbf{Results}: We describe the extension of a previously described non-parametric method to test whether GWAS signals are enriched in functionally defined loci to a situation where only GWAS p values are available.  The approach is implemented in VSEAMS, a freely available software pipeline. We use VSEAMS to integrate functional gene sets defined via transcription factor knock down experiments with GWAS results for type 1 diabetes and find variant set enrichment in gene sets associated with \textit{IKZF3}, \textit{BATF} and  \textit{ESRRA}. \textit{IKZF3} lies in a known T1D susceptibility region, whilst \textit{BATF} and \textit{ESRRA} overlap other immune disease susceptibility regions, validating our approach and suggesting novel avenues of research for type 1 diabetes.\\
\textbf{Availability and implementation}: VSEAMS is available for download (\url{http://github.com/ollyburren/vseams}).\\
\textbf{Contact}: chris.wallace@cimr.cam.ac.uk\\
\doublespacing
\section*{Introduction}

Genome-wide association studies have been successful in identifying  loci associated with many phenotypes \citep{pmid24316577}, and summary statistics  in the form of a list of single SNP p-values for each marker tested,  are increasingly becoming available  in the public domain \citep{pmid20937630,pmid24390342}. In tandem with this, large amounts of functional genomic data across a wide variety of tissues and conditions are increasingly available through public repositories. Methods that integrate genome-wide genetic and genomic data have the potential to provide evidence that functional observations are modulated by underlying genetic variation associated with a particular trait, and are suitable for further study. For example, 50 susceptibility loci are currently described for  type 1 diabetes (\url{http://immunobase.org} accessed 15/03/2014) but the index SNP within only 12 regions  exist as or are in strong linkage disequilibrium (LD) with a non-synonymous coding SNP. This finding agrees with previous research~\citep{pmid22955986,pmid23001124}, and indicates a central role for gene regulatory SNPs in the modulation of complex disease.

One approach is to modify non-parametric approaches developed for microarray pathway analysis \citep{pmid16199517} for use with GWAS study datasets \citep{pmid17966091}. These approaches partner SNPs to genes based on public annotations and then test for differences in evidence of association between two sets of genes, correcting for inter-SNP correlation due to LD. There are several limitations with existing approaches. Firstly classical gene set enrichment analysis is typically based on tests derived from the Kolmogorov-Smirnov, which is under powered and a need for simpler methods has been identified \citep{pmid20048385,pmid23070592}. Secondly, most methods require access to raw genotype data, which are typically not available in the public domain, and such approaches are generally not applicable to meta-analysis based studies. Finally, the permutation based approaches usually employed to adjust for correlation are computationally expensive.  

Nonetheless,  we have previously used a Wilcoxon based method to robustly demonstrate that a human orthologous \textit{IRF7} driven network identified in the rat is enriched for SNPs associated with type 1 diabetes susceptibility \citep{pmid20827270}. In this article, we describe two approximations that allow such tests to be performed with greater computational efficiency and, crucially, without access to raw genotype data. We implement this approach in a freely available software package VSEAMS, and use this to examine enrichment for T1D association among targets of 59 transcription factors identified through knock down experiments in lymphoblastoid cell lines \citep{pmid24603674}. 

\section*{Methods}

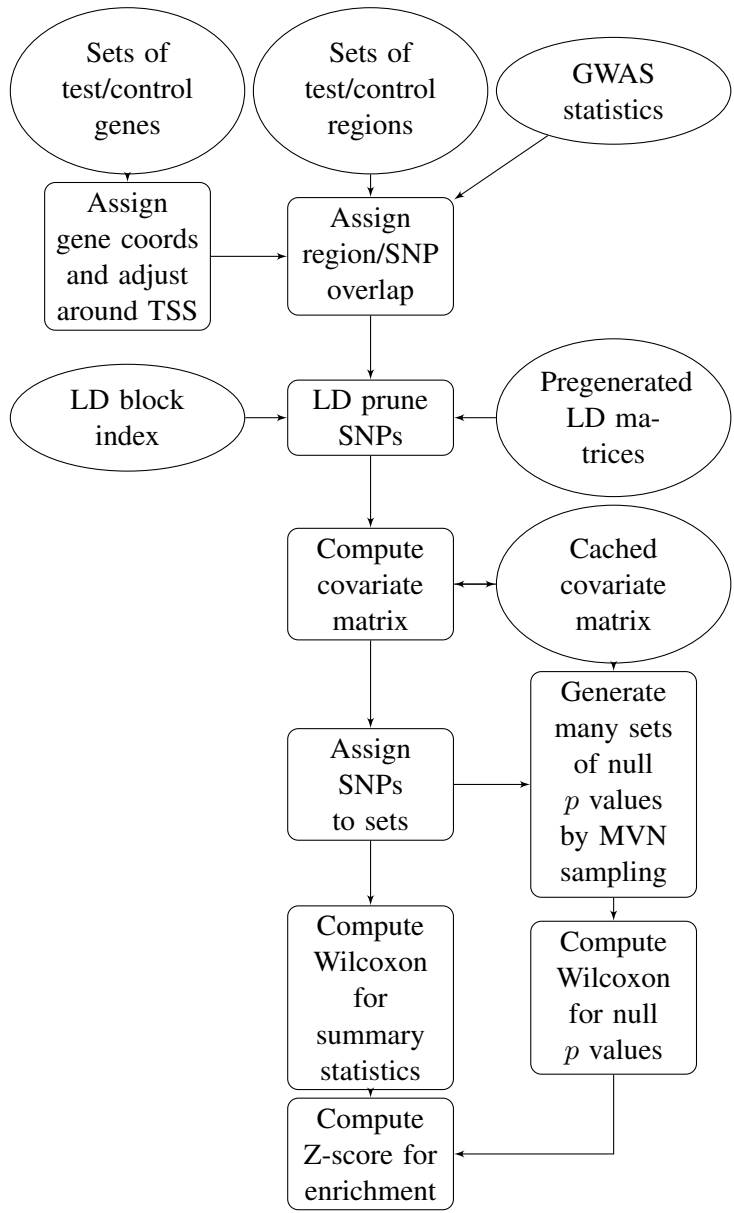
\begin{figure}[!tbp]  
%\centerline{\includegraphics[width=0.5\textwidth,trim=1 100 1 1]{./figure1.pdf}}

 %\begin{tikzpicture}[node distance = 1.5cm,auto]
 \begin{tikzpicture}[auto]
 		\tikzstyle{block} = [rectangle, draw, %fill=blue!20, 
     	text width=5em, text centered, rounded corners, minimum height=1em]
     \tikzstyle{line} = [draw, -latex']
     \tikzstyle{biarrow} = [draw,thick,<>,>=stealth]
 
     \tikzstyle{cloud} = [draw, ellipse,%fill=red!20,
     text width=5em,text centered, minimum height=2em]
     %\tikzstyle{every node}=[font=\tiny]
     % Place nodes
     \matrix [column sep=0.1cm,row sep=0.1cm]
     {        
     %row1
     \node [cloud] (lgenes) {Sets of test/control genes}; &
     \node [cloud] (lregions) {Sets of test/control regions}; &
     \node [cloud] (gwasstats) {GWAS statistics};\\
     %row2
     \node [block] (gcoord) {Assign gene coords and adjust around TSS}; &
     \node [block] (rsnpol) {Assign region/SNP overlap}; &
      \\
     %row3
     \node [cloud] (ldblock) {LD block index}; &
     \node [block] (psnps) {LD prune SNPs}; &
     \node [cloud] (ldmat) {Pregenerated LD matrices};\\
     %row4
     &
     \node [block] (covmat) {Compute covariate matrix};&
     \node [cloud] (cache) {Cached covariate matrix};\\
     %row5
     &
     \node [block] (snpset) {Assign SNPs to sets};&
     \node [block] (perms) {Generate many sets of null $p$ values by MVN sampling};\\
     %row6
     &
     \node [block] (w) {Compute Wilcoxon for summary statistics}; &
     \node [block] (wstar) {Compute Wilcoxon for null $p$ values}; \\
     %row7
     &
     \node [block] (zscore) {Compute Z-score for enrichment}; &
     \\
     };
     % Draw edges
     \path [line] (lregions) -- (rsnpol);
     \path [line] (lgenes) -- (gcoord);
     \path [line] (gwasstats) -- (rsnpol);
     \path [line] (gcoord) -- (rsnpol);
     \path [line] (rsnpol) -- (psnps);
     \path [line] (psnps) -- (covmat);
     \path [line] (covmat) -- (snpset);
     \path [line] (snpset) -- (w);
     \path [line] (w) -- (zscore);
     \path [line] (snpset) -- (perms);
     \path [line] (perms) -- (wstar);
     \path [line] (wstar) |- (zscore);
     \path [line] (cache) -- (covmat);
     \path [line] (covmat) -- (cache);
     \path [line] (ldmat) -- (psnps);
     \path [line] (ldblock) -- (psnps);
     \path [line] (cache) -- (perms);
 \end{tikzpicture}
\caption{The VSEAMS pipeline. VSEAMS takes as input two lists of genes
or regions for comparison, and extracts from GWAS summary statistics
$p$ values for SNPs near those genes or regions.  The observed Wilcoxon rank sum
test statistic is compared to its null distribution determined by its
theoretical mean and a variance derived by simulating null $p$ values
with a correlation structure matching the underlying genotype
structure.  Caching of pregenerated LD matrices reduces computation
time. }
\label{figure1}
\end{figure}

\subsection*{Outline of existing Wilcoxon-based approach}

Given two sets of genes (test and control), our task is to decide whether GWAS association signals for a given trait differ between SNPs near test and control genes.  The assignment of SNPs to genes will be described below.  Once that assignment has been made, we are faced with comparing two distributions of p values.  We use a non-parametric test, the Wilcoxon rank sum test, to test a null hypothesis that these two sets have equal medians.  The test statistic is denoted $W$.  Its mean is known theoretically, but its variance is inflated when SNPs are in any degree of LD. To address this, \citet{pmid20827270} repeatedly permuted case control status in a GWAS dataset to generate replicates of W under the null.  A Z score can then be derived
\begin{equation}
Z = \frac{(W - \mu_0)}{\sqrt{V}},\label{eq:zscore}
\end{equation}
where $W$ is the observed test statistic, $\mu_0$ is its theoretical mean, and $V$ is its empirical variance derived from the replicates of $W$.  Calculation of $V$ is computationally slow, and requires access to the raw genotype data which are not always available.

\subsection*{Creation of LD matrices and indices}
VSEAMS removes the need to access the raw data by instead approximating $V$ by $V^*$, estimated by simulating p values according to the pairwise correlation matrix for a set of SNPs. Given a matrix of $r^2$ values between SNPs of interest, $\boldsymbol{\Sigma}$, which may be derived from public data, we simulate $Z\sim N(0,\boldsymbol{\Sigma})$, from which p values can be derived in the usual manner.  These can be combined to give replicates of $W$, with $V^*$ equal to the empirical variance of these replicates.

Previous software employing this approach - VEGAS \citep{pmid20598278} relies on a set of correlation matrices pregenerated on a predefined gene by gene basis using LD derived from HapMap phase 2 population data \citep{pmid14685227}. To make this approach more flexible to alteration in gene definition and applicable to studies employing SNPs not typed by the HapMap project we developed a methodology to leverage data available from 1000 Genomes Project \citep{pmid20981092}. Computing pairwise LD between all SNPs on a given chromosome is inefficient, therefore using HapMap recombination frequency data we split the genome into contiguous regions of length 0.1cM. We downloaded the EUR 1KG dataset in VCF form (\url{http://www.sph.umich.edu/csg/abecasis/MACH/download/1000G.2012-03-14.html} accessed 02/01/2014) and computed pairwise LD ($r^2$) for each recombination region using \textit{tabix} \citep{pmid21208982} and Bioconductor R libraries \citep{Gentleman2004}. Finally we created an index that allowed fast LD retrieval based on genomic coordinates.

\subsection*{LD pruning GWAS summary statistics}
As described above, LD between SNPs increases the variance of the test statistic. Some LD-based pruning of SNPs in the strongest LD can produce a more stable test statistic. VSEAMS achieves this by using the set of pregenerated LD matrices and hierarchical cluster analysis to select a set of tag SNPs at a user defined $r^2$ threshold. These are then taken forward for analysis. We recommend a relatively relaxed threshold of $r^2 >= 0.95$, removing just those SNPs in extremely strong LD.

\subsection*{Creating SNP sets}

As previously described, VSEAMS uses a non-parametric method which compares the distribution of $p$ values from a GWAS study between two sets of test and control SNPs identified through proximity to test and control genes. The test set, for example, might be the set of genes diffentially expressed in a microarray experiment. As with any competitive test of association, the control set requires careful consideration. For example, one could imagine that genes expressed in lymphocytes are more likely to be related to immune function than their complement. Ignoring this could confound any test of enrichment for association of immune-related phenotypes. Therefore, we encourage users to think carefully about the construction of test and control gene sets, and for microarray derived sets we advocate matching on mean gene expression and coefficient of variation, perhaps using \textit{matching} R package \citep{diamond2012}.                          

The first operation of VSEAMS takes a list of Ensembl \citep{Flicek2013} identifiers for both test and control sets and integrates these with bed-formatted GWAS data to provide a set of test and control SNPs. In order to capture potential regulatory sequences the software allows a user defined offset $\pm$ the transcriptional start site of each gene. Based on \citet{pmid22532805}, which examines the overall distribution of the positions of regulatory SNPs and target genes, we recommend an offset of 200 kb.  For even greater flexibility the software also accepts raw genomic coordinates to define test and control regions sets, and so is not limited to either a single source of annotation or even, gene-centric analysis. In some cases where genes or regions overlap, a variant is assigned to both test and control sets. To allow for this we randomly assign such SNPs to either test or control set. If such overlaps are substantial, we recommend repeating the analysis two or three times to check robustness of any result.

\subsection*{Computing correlation matrices}

We employ a similar method to VEGAS \citep{pmid20598278} to compute correlation matrices. Briefly, using the pregenerated index we identify relevant pregenerated LD matrices, these are then processed to identify the nearest positive definite using Cholesky decomposition implemented in the \textit{corpcor} R package. For efficiency these are computed once and stored as they are applicable for any future analysis using VSEAMS.

\subsection*{Calculation of $V^*$}
The cached correlation matrices are then used to generate multivariate samples of correlated normal variables, $Z$, that mirror the LD-induced correlation in the observed data, using the \textit{mvtnorm} R package~\citep{mvtnormbook}. These are converted to $p$ values in the usual way and, using the R package  \textit{wgsea}, are used to compute replicates of $W$ under the null.  The  empirical variance of these replicates is used to estimate $V^*$.  VSEAMS allows for stratified analysis of multiple GWAS, for example, individual components of a meta analysis study, using van Elteren's method to calculate a combined Z-score \citep{vanelteren60}, although we show below that summary statistics from a meta analysis of multiple GWAS can be used directly.

\subsection*{Prioritisation of genes within enriched gene sets}
Once enrichment is established, VSEAMS can be used to rank the genes based on
summary statistics. For each gene/region in the enriched test set VSEAMS
computes $\bar{P}=mean(-log(p))$ over SNPs assigned to that gene, and using
simulations already available we compute $n$ sets of
$\bar{P}^{*}_{i}=\operatorname{mean}(-\log(p^{*}_{i}))$. An empirical p-value is given by 
$$\frac{1}{n}\sum_{i=1}^n I(\bar{P}^{*}_{i} > \bar{P})$$
where $i$ indexes the $n$ simulated datasets and $I()$ is an indicator function.
Note that as $p \sim U[0,1]$ under a null of no association, $-\log(p) \sim
Exp(1)$, and so $\bar{P}$ is expected to be close to 1 where a gene is not
associated with a given trait.

\subsection*{Type 1 diabetes GWAS datasets}

\citet{pmid19430480} published a meta analysis of three T1D GWAS,
comprising one study using the Affymetrix 500k, ~\citep[WTCCC]{pmid17554300}, and two which used the Illumina
550k chip.  One of these selected cases from Genetics of Kidneys in Diabetes
(GoKinD) study of diabetic nephropathy and reference samples from the National
Institute of Mental Health (NIMH) study and the other used
samples from the Type 1 Diabetes Genetic Consortium ~\citep[T1DGC]{pmid18978792}.  Genotypes were
imputed to allow all SNPs genotyped in either study to be meta analysed in a total of
7,514 cases and 9,045 controls. Due to both its large effect on T1D risk and the
extended LD across the MHC region, we excluded all SNPs in a
window chr6:25Mb..35Mb (GRCh37) from analysis.  We downloaded summary statistics
from T1DBase.org \citep{pmid20937630} and applied quality control thresholds as
described in ~\citet{pmid19430480}.

\subsection*{Validation analyses}
\label{sec:validation-analyses}

Firstly, we wanted to compare the approximate result using summary statistics and $V^*$ to that from direct permutation of the phenotype using $V$.  For this, we used the T1DGC study component for which we have raw genotype data, approximately 4,000 cases and 4,000 controls drawn from the UK population. SNP testing was conducted using the R package \textit{snpStats}.  We selected a random set of 200 protein coding genes (supplementary table \ref{supptab:siglist}) and generated 100 sets of 100 control and 100 test gene sets by randomly sampling from these 200 genes. For each set we computed an enrichment Z-score using, (i) VSEAMS and summary p-values, (ii) permuted case/control status and raw genotype data.  To simulate modest enrichment we repeated these analyses with the $p$ value for each SNP in the test set multiplied by 0.9. In each case, we used 10,000 replicates of $W$ to estimate $V$ and $V^*$.

Secondly, we wanted to confirm that VSEAMS is applicable to meta-analysis.  We generated another set of 1000 control/test gene sets using the method described above.  We computed $p$ values for each set using the meta analysis $p$ values and used 100 replicates of $W$ to estimate $V^*$.

\subsection*{Transcription Factor gene set processing}

\citet{pmid24603674} present the results of differential gene expression in siRNA knock downs of 59 transcription factors and chromatin modifiers in lymphoblastoid cell lines. We downloaded results available in supplemental table 3.  For each transcription factor we created a set of test genes that were differentially expressed at an FDR of 5\%, making sure that the transcription factor itself was excluded from this list, using \textit{qvalues} R package. We created a control set by taking the complement set of genes and removing those with missing values or showing evidence of differential expression at an FDR of  10\%. We ran each test/control set in parallel using VSEAMS, and extended gene regions to incorporate $\pm$200 kb around gene transcriptional start site to best capture regulatory variation \citep{pmid22532805}.  We simulated 100,000 replicates of $W$ to estimate $V^*$.

\section*{Results}

\subsection*{VSEAMS pipeline}
VSEAMS is implemented in R and Perl.  To maximise performance it uses grid based computing and utilises the \textit{macd} queue submission manager. VSEAMS was developed to run using the Sun Grid Engine (SGE) however \textit{macd} is  designed to be extensible to support other high performance computing submission solutions.  All software is available under open source license (GPL v2) from (\url{http://github.com/ollyburren/vseams} and \url{http://github.com/ollyburren/macd}). 

\subsection*{$V^*$ is a good approximation for $V$}

\begin{figure}[!tbp]
\centerline{\includegraphics[width=0.5\textwidth]{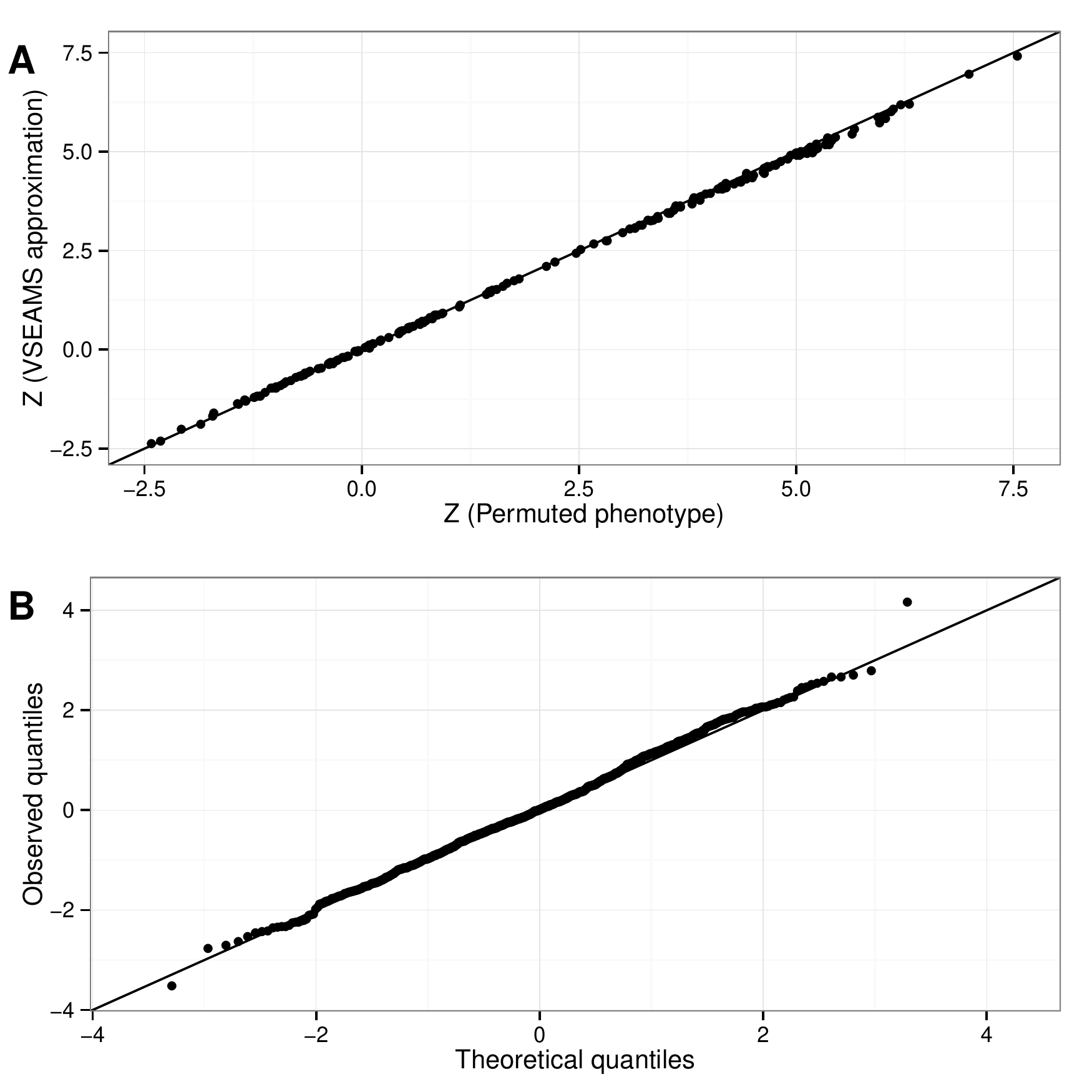}}
\caption{Comparison of Z-score statistics generated by VSEAMS using a simulated set of 200 randomly assigned genes. \textbf{Panel A} compares Z scores from using permuted phenotype vs using summary P-values and VSEAMS (10,000 permutations) for T1DGC study, over 100 gene sets. \textbf{Panel B} shows a qq plot using VSEAMS (100 permutations) applied to the meta analysis of \citet{pmid19430480} for 1000 gene sets.} 
\label{fig:simulation}                    
\end{figure}

Figure \ref{fig:simulation} shows there is good correlation between results obtained from VSEAMS approximations and those from directly permuting genotype (panel \textbf{A}). We also found that $Z$ scores calculated by our approximate method showed a close fit to their theoretical distribution (panel \textbf{B}).  Taken together these results indicate that VSEAMS is a suitable replacement where raw genotyping data are not available and is applicable in the case of a meta-analysis which may include both imputation and different genotyping platforms.

\subsection*{Type 1 diabetes susceptibility enrichment in targets of the  transcription factors \textit{IKZF3}, \textit{BATF} and \textit{ESRRA}}
\begin{figure*}[!tbp]
\centerline{\includegraphics[width=\textwidth]{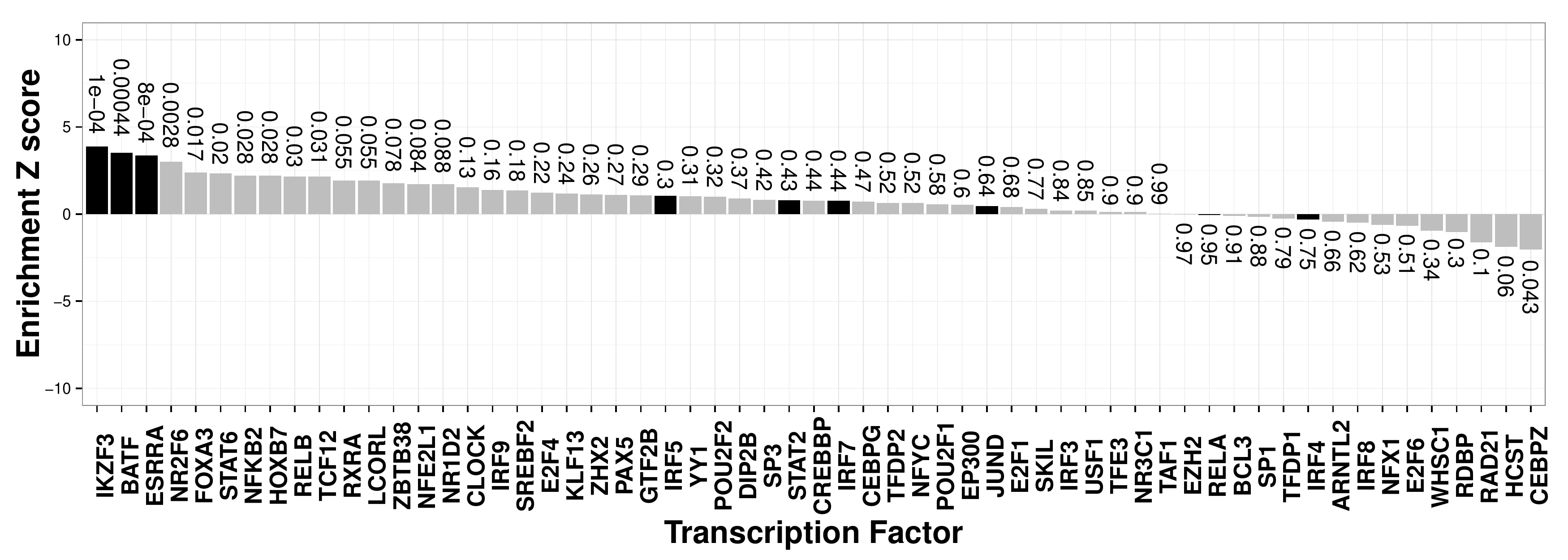}}
\caption{Type 1 diabetes susceptible SNP enrichment (excluding MHC) within transcription factor perturbed gene sets from \citet{pmid24603674} SNPs are pruned on the basis of $r^2$ threshold $\geq$0.95. A positive Z score indicates enrichment, labels denote associated p values. Black bars indicate that the knocked down transcription factor overlaps a known autoimmune susceptibility locus curated in ImmunoBase.}
\label{fig:results_bar}
\end{figure*}

We investigated whether any of the sets of genes perturbed by 59 knock down experiments presented by \citet{pmid24603674} were enriched for SNPs associated with type 1 diabetes susceptibility (figure \ref{fig:results_bar}). Three factors reached significance after Bonferonni correction: \textit{IKZF3} ($p=1.1$ x $10^{-4}, n=1798$), \textit{BATF} ($p=4.4$ x $10^{-4},n=210$) and \textit{ESRRA} ($p=8.0$ x $10^{-4},n=614$), where $n$ is the number of genes in each set. Fourteen genes are common to all three sets (supplementary figure \ref{suppfig:venn} and supplementary table \ref{supptab:commongenes}).

We used VSEAMS to prioritise individual genes within each significant set, selecting 95 out of 2,326 that exceeded Bonferonni threshold for that set (supplementary \ref{supptab:siglist}). Of these, 63 overlap regions of known  type 1 diabetes susceptibility (\url{http://immunobase.org} accessed 15/03/2014).  We draw attention to ten genes that have no conclusively established association to type 1 diabetes but have been highlighted for other immune-modulated diseases in ImmunoBase (table \ref{tab:sig_genes}), three of which are implicated as candidate causal genes in one or more diseases \textit{TRAF3IP2} in psoriasis, ulcerative colitis and Crohn's disease \citep{pmid23143594,pmid23128233}, \textit{ZNF438} in multiple sclerosis \citep{pmid21833088} and \textit{RUNX3} in ankylosing spondylitis and psoriasis \citep{pmid23749187,pmid23143594}.

\begin{table*}[!p]

\caption{Genes with significant gene prioritisation statistics identified from enriched gene sets not overlapping known T1D susceptibility regions. `Disease overlaps' indicates that the interval defined overlaps a disease annotated in {{http://immunobase.org}}. Ankylosing spondylitis (AS), celiac disease (CEL), Crohn's disease (CRO), juvenile idiopathic arthritis (JIA), multiple sclerosis (MS), psoriasis (PSO), rheumatoid arthritis (RA), ulcerative colitis (UC). Coordinates are given for build GRCh37. $^a$Gene is implicated as causal in that disease. $^b$Regions overlap. \label{tab:sig_genes}}
\begin{tabular}{p{2cm}p{3cm}p{1.5cm}p{1.5cm}p{4.5cm}p{3cm}}
  \hline
  Transcription Factor & Ensembl ID & HGNC Symbol & $P$ (empirical) & Coordinates & Disease Overlap \\\hline
  \textit{IKZF3} &	ENSG00000056972 &	\textit{TRAF3IP2} &	$< 10^{-6}$ &	chr6:111727481..112127481 & CRO$^a$, PSO$^a$ ,UC$^a$ \\
  \textit{IKZF3} &	ENSG00000183621 &	 \textit{ZNF438} &	0.000008 & chr10:31109136..31520866 & MS$^a$, RA \\
  \textit{IKZF3} &	ENSG00000110344	 & 	\textit{UBE4A} & 		$< 10^{-6}$	 & 	chr11:118030300..118430300 & CEL, MS, PBC, RA, SJO \\
  \textit{IKZF3} &	ENSG00000108465 &	\textit{CDK5RAP3} &	0.000003 &	chr17:45845176..46245176 & AS, MS\\
  \textit{IKZF3} & 	ENSG00000105655	 & 	\textit{ISYNA1} & 		0.000006 & 		chr19:18349111..18749111 & MS \\
  \textit{IKZF3}	 & 	ENSG00000128268 & 		\textit{MGAT3} & 		0.000004 & 		chr22:39653349..40053349 & CD, PBC, UC\\
  \textit{BATF} & 		ENSG00000020633 & 		\textit{RUNX3} & 		0.000169 & 		chr1:25091612..25491612 & AS$^a$, PS$^a$\\
  \textit{BATF} & 		ENSG00000241685 &\textit{	ARPC1A}	 & 	0.000218 & 		chr7:98723521..99123521 & CD, UC \\
  \textit{ESRRA} & 		ENSG00000213619 & 		\textit{NDUFS3} & 		0.000051 &	chr11:47386888..47786888$^b$ & MS \\
  \textit{ESRRA} & 		ENSG00000123444 & 		\textit{KBTBD4} & 		0.000082 & 		chr11:47400567..47800567$^b$ & MS \\
 
\end{tabular}{}

\end{table*}

\section*{Discussion}

There are  caveats when inferring observations between this
and cell types systems that are important in type 1 diabetes aetiology. However,
the three transcription factors we identify have been previously implicated in
autoimmunity when cross referenced with data from ImmunoBase
(\url{http://immunobase.org} accessed 03/04/2014).  
\textit{IKZF3} is a transcription factor located within a type 1 diabetes
susceptibility region at 17q12 \citep{pmid19430480}  and overlaps susceptibility loci for
ulcerative colitis, Crohn's disease, primary billiary cirrhosis, and rheumatoid
arthritis \citep{pmid23128233,pmid22961000,pmid20453842}.
\textit{IKZF3} is
implicated in the regulation of B cell lymphocyte proliferation and
differentiation \citep{pmid9155026}. \textit{BATF}  overlaps rheumatoid arthritis and
multiple sclerosis susceptibility loci at 14q24.3 \citep{pmid20453842,pmid21833088}. Mice over expressing
\textit{Batf} show impaired T-cell development in vitro and no induction of IL-2
\citep{pmid12594265}. \textit{ESRRA} overlaps alopecia areate, Crohn's disease,
multiple sclerosis and ulcerative colitis loci at 11q13.1 \citep{pmid20596022,pmid23128233,pmid21833088} and is a metabolic
regulator of T-cell activation and differentiation \citep{pmid22042850}. 
Future work will determine whether the
enrichment pattern observed with type 1 diabetes is shared with, or distinct from, other autoimmune
traits.

It is of note that the set of genes perturbed when IRF7 is knocked down shows no
evidence for enrichment, in contrast to our previous work. This is likely to
reflect the fact that  the transcription factor experiments were performed in a
lymphoblastoid cell line. The master regulator  of the IRF7 network previously
described is \textit{GPR183}, and is known to be activated by exposure to
Epstein-Barr virus, therefore IRF7 responsiveness is likely to be altered
\citep{pmid20827270} in LCL's, which emphasises a need for transcription factor
function to be studied in primary cells.  

Correlation is a problem for all enrichment analyses because it results in inflated test statistics compared to their theoretical distribution.  This problem exists in gene set enrichment analyses (GSEA) of gene expression expression datasets, but is more pronounced for SNP data, in which historical recombination events produce LD patterns that are both complex and strong.  The original GSEA method accounts for this correlation by permuting phenotypes and repeating the entire gene expression analysis multiple times \citep{pmid16199517}, an approach we also took in a previous variant set enrichment analysis \citep{pmid20827270}. This computationally intensive approach seems required because permuting SNPs or genes directly destroys the correlation structure.  Tests have been adapted for gene set enrichment that deal theoretically with the inflation of variance by estimating an average variance inflation factor \citep{pmid22638577}, but for SNPs we do not believe a single variance inflation factor can capture the strength and highly variable correlation observed.  Instead, in VSEAMS, we adapt a multivariate normal sampling approach  which is not only faster than phenotype permutation, but allows application in the typical case where raw genotype data are not available.  Although this framework could equally be applied to parametric tests such as t-tests, we chose to concentrate on a non-parametric test because it is more robust to occasional genotyping errors which may arise and which, without access to genotyping data, are impossible to check.

Although the selection of test sets is often straightforward, the selection of proper control sets tends not to be.  This requires careful understanding of the competitive hypothesis tested in enrichment studies and consideration needs to be made, for example, when gene sets are derived from raw differential expression analysis that the control set is selected from a matched distribution of non differentially expressed genes, to prevent confounding.  Here, we restricted our set of control genes to genes that were perturbed by at least one transcription factor in the  lymphoblastoid cell line knock down experiments \citep{pmid24603674}.

Imprecise knowledge of regulatory variants for individual genes hampers any test of variant set enrichment.  As regulatory variation may lie 200kb from a gene \citep{pmid22532805}, we use a large window to assign SNPs to genes.  This increases the likelihood of overlapping regions occurring in test and control sets. We have implemented a random assignment strategy to mitigate this, and, although unbiased, this approach can result in a loss of power in the test for enrichment.  Combination of chromatin state annotation with high-throughput chromatin conformation capture (``Hi-C''), has the potential to allow better definition of genomic regions involved in regulating specific genes.  This increased resolution will require a corresponding increase in GWAS resolution through the use of imputation.  Additionally, as regulatory function varies in a cell specific manner, annotation of multiple primary cell types and careful consideration of the biologically relevant cell types will be required. However, we expect this more precise definition of functional SNP sets will allow a sharp increase in the power of variant set enrichment analyses which will allow VSEAMS analyses to interpret functionally defined genetic regions by linking them to end point phenotypes.

\singlespacing
\section*{Acknowledgements}
We thank John Todd for his help in conceiving the study, interpreting the results and comments on the manuscript. We thank Vin Everett and Wojciech Geil for computing support, as well as other members of the Diabetes and Inflammation laboratory for assistance throughout. We acknowledge Darren Cusanovich for facilitating early access to knockdown experimental data. 
                                                               
This study utilises resources provided by the Type 1 Diabetes Genetics Consortium, a collaborative clinical study sponsored by the National Institute of Diabetes and Digestive and Kidney Diseases, National Institute of Allergy and Infectious Diseases, National Human Genome Research Institute, National Institute of Child Health and Human Development, and Juvenile Diabetes Research Foundation International and supported by {[}U01 DK062418{]}. This study makes use of data generated by the Wellcome Trust Case Control Consortium. A full list of the investigators who contributed to the generation of the data are available from {{http://www.wtccc.org.uk/}}. Funding for the project was provided by the Wellcome Trust under award {[}076113{]}. We acknowledge the National Institute of Mental Health for Control subjects from the National Institute of Mental Health Schizophrenia Genetics Initiative (NIMH-GI), data and biomaterials are being collected by the `Molecular Genetics of Schizophrenia II' collaboration. The investigators and co-investigators are as follows: P.V. Gejman (Collaboration coordinator) and A.R. Sanders (ENH/Northwestern University, {[}MH059571{]}); F. Amin (Emory University School of Medicine, {[}MH59587{]}); N. Buccola (Louisiana State University Health Sciences Center, {[}MH067257{]}); W. Byerley (University of California-Irvine, {[}MH60870{]}); C.R. Cloninger (Washington University, St. Louis, U01, {[}MH060879{]}); R. Crowe (PI) and D. Black (University of Iowa, {[}MH59566{]}); R. Freedman (University of Colorado, {[}MH059565{]}); D. Levinson (University of Pennsylvania, {[}MH061675{]}); B. Mowry (University of Queensland, {[}MH059588{]}); and J. Silverman (Mt. Sinai School of Medicine, {[}MH59586{]}). The samples were collected by V.L. Nimgaonkar's group at the University of Pittsburgh as part of a multi-institutional collaborative research project with J. Smoller and P. Sklar (Massachusetts General Hospital, {[}MH 63420{]}).We gratefully acknowledge the Genetics of Kidneys in Diabetes (GoKinD) study obtained from the Genetic Association Information Network (GAIN) database found at {\url{http://view.ncbi.nlm.nih.gov/dbgap/} through dbGaP accession number phs000018.v1.p1 

%\paragraph{Funding\textcolon}
Funding:\\
This work was funded by the JDRF{[}9-2011-253{]}, the Wellcome Trust {[}091157{]}, the National Institute for Health Research Cambridge Biomedical Research Centre and  the European Union’s 7th Framework Programme {[}FP7/2007-2013{]} under grant agreement {[}241447{]}. The Cambridge Institute for Medical Research is in receipt of a Wellcome Trust Strategic Award {[}100140{]}. Chris Wallace and Hui Guo are supported by the Wellcome Trust {[}089989{]}. ImmunoBase.org is supported by Eli Lilly and Company.

\bibliographystyle{bioinformatics}
\bibliography{burren_et_al_vseams}

\subsection*{Supplementary tables and figures}
\setcounter{figure}{0}
\setcounter{table}{0}
\renewcommand{\tablename}{Supplementary Table }
\renewcommand{\figurename}{Supplementary Figure } 
                                                               
%\begin{figure}[ht]
%\includegraphics[width=0.9\textwidth,height=0.9\textheight,keepaspectratio]{./figures/figure6_supplemental.pdf}
%\caption{Plot showing how variance of $W^*$ varies with number of permutations in transcription factor gene sets showing enrichment. The first 500 permutations have been ommitted for clarity.}
%\label{suppfig:tfvarw}
%\end{figure}

\begin{figure}[ht]
\includegraphics[width=0.9\textwidth,height=0.9\textheight,keepaspectratio]{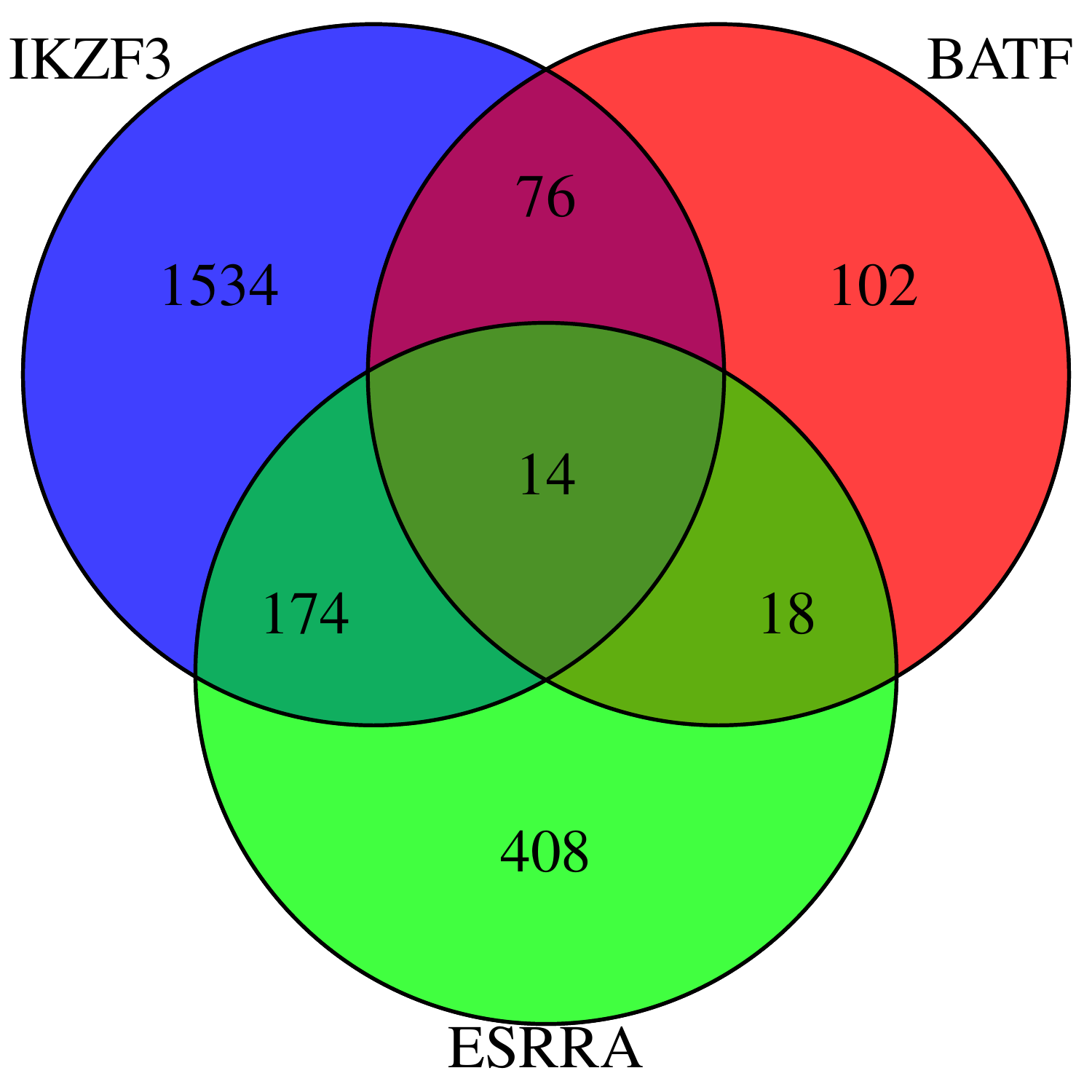}
\caption{Overlap between the different genesets showing significant enrichment for type 1 diabetes association from \textit{Cusanovich et al.}, \textit{IKZF3}(n=1798), \textit{BATF}(n=210) and \textit{ESRAA}(n=614).}
\label{suppfig:venn}
\end{figure}

\begin{table}[ht]

\begin{tabular}{llcrc}\hline  
 Ensembl ID & HGNC Symbol & Strand & Coordinates & Disease Overlap \\\hline     
	ENSG00000116574 & RHOU &  + & chr1:228870824..228882416 & \\ 
	ENSG00000121957 & GPSM2 &  + & chr1:109417972..109477167 & \\         
	ENSG00000188404 & SELL & - & chr1:169659808..169680839 & \\ 
	ENSG00000151694 & ADAM17 & - & chr2:9628615..9695921 & \\ 
	ENSG00000085719 & CPNE3 &  + & chr8:87497059..87573726 & \\ 
	ENSG00000136982 & DSCC1 & - & chr8:120846216..120868250 & \\ 
	ENSG00000107223 & EDF1 & - & chr9:139756571..139760738 & \\ 
	ENSG00000187742 & SECISBP2 &  + & chr9:91933421..91974557 & \\ 
	ENSG00000149289 & ZC3H12C &  + & chr11:109964087..110042566 & PSO \\ 
	ENSG00000181019 & NQO1 & - & chr16:69740899..69760854 & \\ 
	ENSG00000198417 & MT1F &  + & chr16:56691606..56694610 & \\ 
	ENSG00000224161 & RPS26P54 &  + & chr18:57428790..57429137 & \\ 
	ENSG00000160190 & SLC37A1 &  + & chr21:43916118..44001550 & \\ 
	ENSG00000102393 & GLA & - & chrX:100652791..100662913 & \\           
   \hline
\end{tabular}
\caption{14 Genes overlapping all 3 transcription factor knockdown genesets enriched for T1D associated variants, Psoriasis (PSO)}
 \label{supptab:commongenes} 
\end{table}           

%\begin{table}[ht]
%\centering
\afterpage{
\LTcapwidth=\textwidth
\begin{longtable}{r|l|r|r|l|l|l}
    \hline
 Number & Chromosome & Start & End & Strand & EnsemblID & HGNCID \\ 
  \hline
  \endfirsthead
  
  \multicolumn{7}{c}%
  {{\bfseries \tablename\ \thetable{} -- continued from previous page}} \\
  \hline
 Number & Chromosome & Start & End & Strand & EnsemblID & HGNCID \\ 
  \hline
  \endhead
  
  \hline \multicolumn{7}{r}{{Continued on next page}} \\ 
  \endfoot
  
  \hline \hline
  \endlastfoot
  
  1 & chr1 & 8412457 & 9077702 & - & ENSG00000142599 & RERE \\ 
  2 & chr1 & 10290159 & 10690159 & + & ENSG00000175279 & APITD1 \\ 
 3 & chr1 & 13190765 & 13590765 & - & ENSG00000182330 & PRAMEF8 \\ 
 4 & chr1 & 15711605 & 16111605 & - & ENSG00000116771 & AGMAT \\ 
 5 & chr1 & 25464408 & 25864408 & + & ENSG00000183726 & TMEM50A \\ 
 6 & chr1 & 32487529 & 32887529 & + & ENSG00000084623 & EIF3I \\ 
 7 & chr1 & 32617122 & 33017122 & + & ENSG00000162526 & TSSK3 \\ 
 8 & chr1 & 33346597 & 33746597 & - & ENSG00000004455 & AK2 \\ 
 9 & chr1 & 40742887 & 41142887 & + & ENSG00000187815 & ZFP69 \\ 
 10 & chr1 & 92214928 & 92614928 & + & ENSG00000137948 & BRDT \\ 
 11 & chr1 & 93445476 & 93845476 & + & ENSG00000122483 & CCDC18 \\ 
 12 & chr1 & 109958726 & 110358726 & + & ENSG00000116337 & AMPD2 \\ 
 13 & chr1 & 110750564 & 111150564 & - & ENSG00000134248 & LAMTOR5 \\ 
 14 & chr1 & 112809163 & 113209163 & + & ENSG00000134245 & WNT2B \\ 
 15 & chr1 & 114853781 & 115253781 & - & ENSG00000197323 & TRIM33 \\ 
 16 & chr1 & 145089772 & 145489772 & + & ENSG00000163386 & NBPF10 \\ 
 17 & chr1 & 151145209 & 151545209 & - & ENSG00000143416 & SELENBP1 \\ 
 18 & chr1 & 153431130 & 153831130 & + & ENSG00000143553 & SNAPIN \\ 
 19 & chr1 & 153695451 & 154095451 & - & ENSG00000143614 & GATAD2B \\ 
 20 & chr1 & 157322310 & 157722310 & - & ENSG00000143297 & FCRL5 \\ 
 21 & chr1 & 178850512 & 179250512 & + & ENSG00000186283 & TOR3A \\ 
 22 & chr1 & 206943970 & 207343970 & - & ENSG00000162897 & FCAMR \\ 
 23 & chr1 & 211231719 & 211631719 & + & ENSG00000117625 & RCOR3 \\ 
 24 & chr1 & 212258879 & 212658879 & + & ENSG00000066027 & PPP2R5A \\ 
 25 & chr1 & 223366715 & 223766715 & + & ENSG00000178395 & C1orf65 \\ 
 26 & chr2 & 105682585 & 106082585 & - & ENSG00000268809 &  \\ 
 27 & chr2 & 113322254 & 113722254 & - & ENSG00000169607 & CKAP2L \\ 
 28 & chr2 & 127976003 & 128376003 & + & ENSG00000115718 & PROC \\ 
 29 & chr2 & 130739310 & 131139310 & + & ENSG00000152082 & MZT2B \\ 
 30 & chr2 & 171847333 & 172287824 & - & ENSG00000198586 & TLK1 \\ 
 31 & chr2 & 219997899 & 220397899 & - & ENSG00000182698 & RESP18 \\ 
 32 & chr3 & 32233163 & 32633163 & + & ENSG00000153551 & CMTM7 \\ 
 33 & chr3 & 42747223 & 43147223 & + & ENSG00000182983 & ZNF662 \\ 
 34 & chr3 & 45727996 & 46127996 & + & ENSG00000173585 & CCR9 \\ 
 35 & chr3 & 49767606 & 50167606 & - & ENSG00000164077 & MON1A \\ 
 36 & chr3 & 111497857 & 111897857 & + & ENSG00000144827 & ABHD10 \\ 
 37 & chr3 & 124798021 & 125198021 & - & ENSG00000221955 & SLC12A8 \\ 
 38 & chr3 & 125039041 & 125439041 & - & ENSG00000114520 & SNX4 \\ 
 39 & chr3 & 125455882 & 125855882 & - & ENSG00000189366 & ALG1L \\ 
 40 & chr3 & 133571028 & 133971028 & - & ENSG00000174640 & SLCO2A1 \\ 
 41 & chr3 & 145768966 & 146168966 & - & ENSG00000114698 & PLSCR4 \\ 
 42 & chr3 & 148383043 & 148783043 & + & ENSG00000163751 & CPA3 \\ 
 43 & chr3 & 183779251 & 184179251 & - & ENSG00000163888 & CAMK2N2 \\ 
 44 & chr3 & 189840264 & 190240264 & - & ENSG00000163347 & CLDN1 \\ 
 45 & chr4 & 6517842 & 6917842 & + & ENSG00000186222 & BLOC1S4 \\ 
 46 & chr4 & 9145874 & 9545874 & + & ENSG00000235780 & USP17L27 \\ 
 47 & chr4 & 38984024 & 39384024 & + & ENSG00000157796 & WDR19 \\ 
 48 & chr4 & 44175926 & 44650824 & - & ENSG00000183783 & KCTD8 \\ 
 49 & chr4 & 71568043 & 71968043 & + & ENSG00000173542 & MOB1B \\ 
 50 & chr4 & 140022609 & 140422609 & + & ENSG00000164134 & NAA15 \\ 
 51 & chr5 & 133140824 & 133540824 & - & ENSG00000213585 & VDAC1 \\ 
 52 & chr5 & 137678989 & 138078989 & - & ENSG00000120705 & ETF1 \\ 
 53 & chr5 & 139287362 & 139687362 & + & ENSG00000185129 & PURA \\ 
 54 & chr5 & 140007563 & 140407563 & + & ENSG00000081842 & PCDHA6 \\ 
 55 & chr5 & 176126333 & 176526333 & - & ENSG00000160883 & HK3 \\ 
 56 & chr5 & 176360026 & 176760026 & + & ENSG00000165671 & NSD1 \\ 
 57 & chr5 & 176578853 & 176978853 & - & ENSG00000169223 & LMAN2 \\ 
 58 & chr5 & 177357997 & 177757997 & + & ENSG00000145916 & RMND5B \\ 
 59 & chr6 & 4506393 & 4955785 & + & ENSG00000153046 & CDYL \\ 
 60 & chr6 & 25833796 & 26233796 & - & ENSG00000137259 & HIST1H2AB \\ 
 61 & chr6 & 29490552 & 29890552 & + & ENSG00000204642 & HLA-F \\ 
 62 & chr6 & 32843703 & 33243703 & + & ENSG00000223865 & HLA-DPB1 \\ 
 63 & chr6 & 79587953 & 79987953 & - & ENSG00000146247 & PHIP \\ 
 64 & chr6 & 91096764 & 91496764 & - & ENSG00000135341 & MAP3K7 \\ 
 65 & chr6 & 111103218 & 111503218 & + & ENSG00000197498 & RPF2 \\ 
 66 & chr6 & 116666773 & 117066773 & - & ENSG00000173626 & TRAPPC3L \\ 
 67 & chr6 & 130134844 & 130534844 & + & ENSG00000198945 & L3MBTL3 \\ 
 68 & chr6 & 131256806 & 131656806 & + & ENSG00000118507 & AKAP7 \\ 
 69 & chr6 & 132884598 & 133284598 & - & ENSG00000112303 & VNN2 \\ 
 70 & chr6 & 139413276 & 139813276 & - & ENSG00000164440 & TXLNB \\ 
 71 & chr6 & 142209936 & 142609936 & - & ENSG00000135577 & NMBR \\ 
 72 & chr6 & 153352455 & 153752455 & + & ENSG00000213121 &  \\ 
 73 & chr6 & 165523096 & 165923096 & - & ENSG00000112539 & C6orf118 \\ 
 74 & chr7 & 5265045 & 5665045 & - & ENSG00000182095 & TNRC18 \\ 
 75 & chr7 & 6188612 & 6588612 & - & ENSG00000178397 & FAM220A \\ 
 76 & chr7 & 21382833 & 21941457 & + & ENSG00000105877 & DNAH11 \\ 
 77 & chr7 & 30123923 & 30523923 & + & ENSG00000180233 & ZNRF2 \\ 
 78 & chr7 & 32733743 & 33133743 & - & ENSG00000170852 & KBTBD2 \\ 
 79 & chr7 & 35535181 & 35935181 & - & ENSG00000122557 & HERPUD2 \\ 
 80 & chr7 & 36893961 & 37688852 & - & ENSG00000155849 & ELMO1 \\ 
 81 & chr7 & 44380914 & 44780914 & - & ENSG00000015520 & NPC1L1 \\ 
 82 & chr7 & 44421886 & 44821886 & - & ENSG00000158604 & TMED4 \\ 
 83 & chr7 & 47928225 & 48328225 & + & ENSG00000183696 & UPP1 \\ 
 84 & chr7 & 72897783 & 73297783 & - & ENSG00000176410 & DNAJC30 \\ 
 85 & chr7 & 89813035 & 90213035 & + & ENSG00000157224 & CLDN12 \\ 
 86 & chr7 & 107184142 & 107584142 & + & ENSG00000105879 & CBLL1 \\ 
 87 & chr7 & 107243670 & 107643670 & - & ENSG00000091138 & SLC26A3 \\ 
 88 & chr7 & 108010110 & 108410110 & - & ENSG00000135241 & PNPLA8 \\ 
 89 & chr7 & 127801739 & 128201739 & - & ENSG00000224940 & PRRT4 \\ 
 90 & chr7 & 140190577 & 140590577 & + & ENSG00000090266 & NDUFB2 \\ 
 91 & chr8 & 6595860 & 6995860 & - & ENSG00000164821 & DEFA4 \\ 
 92 & chr8 & 6637602 & 7037602 & - & ENSG00000206047 & DEFA1 \\ 
 93 & chr8 & 22235792 & 22635792 & + & ENSG00000120913 & PDLIM2 \\ 
 94 & chr8 & 63881112 & 64281112 & + & ENSG00000185728 & YTHDF3 \\ 
 95 & chr8 & 94567072 & 94967072 & + & ENSG00000164953 & TMEM67 \\ 
 96 & chr8 & 110146614 & 110546614 & - & ENSG00000120526 & NUDCD1 \\ 
 97 & chr8 & 110788076 & 111188076 & - & ENSG00000164794 & KCNV1 \\ 
 98 & chr8 & 117687105 & 118087105 & - & ENSG00000164754 & RAD21 \\ 
 99 & chr8 & 133572958 & 133972958 & - & ENSG00000165071 & TMEM71 \\ 
 100 & chr8 & 141445718 & 141845718 & - & ENSG00000123908 & AGO2 \\ 
 101 & chr9 & 15311017 & 15711017 & - & ENSG00000164985 & PSIP1 \\ 
 102 & chr9 & 19030433 & 19430433 & + & ENSG00000137145 & DENND4C \\ 
 103 & chr9 & 46189110 & 46589110 & + & ENSG00000231997 & FAM27D1 \\ 
 104 & chr9 & 114942217 & 115342217 & + & ENSG00000119471 & HSDL2 \\ 
 105 & chr9 & 130287152 & 130687152 & - & ENSG00000187024 & PTRH1 \\ 
 106 & chr9 & 138170925 & 138570925 & + & ENSG00000196422 & PPP1R26 \\ 
 107 & chr9 & 138331386 & 138731386 & - & ENSG00000204007 & GLT6D1 \\ 
 108 & chr9 & 139068133 & 139468133 & - & ENSG00000187796 & CARD9 \\ 
 109 & chr9 & 139765040 & 140165040 & - & ENSG00000186193 & SAPCD2 \\ 
 110 & chr10 & 37947034 & 38347034 & - & ENSG00000198105 & ZNF248 \\ 
 111 & chr10 & 70739988 & 71139988 & + & ENSG00000156502 & SUPV3L1 \\ 
 112 & chr10 & 73411126 & 73811126 & - & ENSG00000197746 & PSAP \\ 
 113 & chr10 & 74812451 & 75212451 & - & ENSG00000182180 & MRPS16 \\ 
 114 & chr10 & 112431565 & 112831565 & + & ENSG00000150593 & PDCD4 \\ 
 115 & chr11 & 724894 & 1124894 & + & ENSG00000183020 & AP2A2 \\ 
 116 & chr11 & 5928914 & 6328914 & + & ENSG00000180919 & OR56B4 \\ 
 117 & chr11 & 31331297 & 31805546 & + & ENSG00000109911 & ELP4 \\ 
 118 & chr11 & 47387121 & 47787121 & - & ENSG00000149187 & CELF1 \\ 
 119 & chr11 & 48146472 & 48546472 & + & ENSG00000176547 & OR4C3 \\ 
 120 & chr11 & 55705194 & 56105194 & - & ENSG00000167822 & OR8J3 \\ 
 121 & chr11 & 72185635 & 72585635 & - & ENSG00000186642 & PDE2A \\ 
 122 & chr11 & 89243467 & 89643467 & + & ENSG00000214414 & TRIM77 \\ 
 123 & chr11 & 95709762 & 96276344 & - & ENSG00000184384 & MAML2 \\ 
 124 & chr11 & 111549659 & 111949659 & + & ENSG00000137720 & C11orf1 \\ 
 125 & chr11 & 111757497 & 112157497 & + & ENSG00000204370 & SDHD \\ 
 126 & chr11 & 128575930 & 128975930 & - & ENSG00000174370 & C11orf45 \\ 
 127 & chr12 & 899219 & 1299219 & - & ENSG00000002016 & RAD52 \\ 
 128 & chr12 & 4558213 & 4958213 & - & ENSG00000111254 & AKAP3 \\ 
 129 & chr12 & 10124737 & 10524737 & - & ENSG00000173391 & OLR1 \\ 
 130 & chr12 & 21454603 & 21854603 & - & ENSG00000004700 & RECQL \\ 
 131 & chr12 & 55620038 & 56020038 & + & ENSG00000185821 & OR6C76 \\ 
 132 & chr12 & 56839798 & 57239798 & - & ENSG00000110955 & ATP5B \\ 
 133 & chr12 & 89547048 & 89947048 & - & ENSG00000139318 & DUSP6 \\ 
 134 & chr12 & 105180088 & 105580088 & + & ENSG00000151131 & C12orf45 \\ 
 135 & chr12 & 105429068 & 105829068 & + & ENSG00000235162 & C12orf75 \\ 
 136 & chr12 & 120439038 & 120839038 & - & ENSG00000089157 & RPLP0 \\ 
 137 & chr12 & 123001439 & 123401439 & - & ENSG00000255398 & HCAR3 \\ 
 138 & chr13 & 31280328 & 31680328 & + & ENSG00000102802 & MEDAG \\ 
 139 & chr13 & 41506882 & 41906882 & - & ENSG00000165572 & KBTBD6 \\ 
 140 & chr13 & 49622047 & 50022047 & + & ENSG00000102543 & CDADC1 \\ 
 141 & chr13 & 52178293 & 52578293 & - & ENSG00000102796 & DHRS12 \\ 
 142 & chr13 & 53026844 & 53426844 & + & ENSG00000165416 & SUGT1 \\ 
 143 & chr13 & 77260540 & 77660540 & - & ENSG00000178695 & KCTD12 \\ 
 144 & chr14 & 24700160 & 25100160 & - & ENSG00000139899 & CBLN3 \\ 
 145 & chr14 & 77382911 & 77782911 & + & ENSG00000165548 & TMEM63C \\ 
 146 & chr14 & 94396590 & 94796590 & - & ENSG00000119632 & IFI27L2 \\ 
 147 & chr14 & 102858998 & 103258998 & + & ENSG00000089902 & RCOR1 \\ 
 148 & chr14 & 104846021 & 105246021 & + & ENSG00000184601 & C14orf180 \\ 
 149 & chr15 & 27016429 & 27778373 & + & ENSG00000182256 & GABRG3 \\ 
 150 & chr15 & 28929629 & 29410518 & + & ENSG00000034053 & APBA2 \\ 
 151 & chr15 & 40026347 & 40426347 & + & ENSG00000128829 & EIF2AK4 \\ 
 152 & chr15 & 40374787 & 40774787 & - & ENSG00000230778 & ANKRD63 \\ 
 153 & chr15 & 59239899 & 59639899 & + & ENSG00000268327 &  \\ 
 154 & chr15 & 89564982 & 89964982 & - & ENSG00000140522 & RLBP1 \\ 
 155 & chr16 & -72994 & 327006 & + & ENSG00000103152 & MPG \\ 
 156 & chr16 & 237113 & 637113 & - & ENSG00000129925 & TMEM8A \\ 
 157 & chr16 & 21818959 & 22218959 & + & ENSG00000185716 & C16orf52 \\ 
 158 & chr16 & 29266285 & 29666285 & - & ENSG00000183336 & BOLA2 \\ 
 159 & chr16 & 30917428 & 31317428 & + & ENSG00000103507 & BCKDK \\ 
 160 & chr16 & 81572702 & 81991899 & + & ENSG00000197943 & PLCG2 \\ 
 161 & chr17 & 2296504 & 2696504 & + & ENSG00000007168 & PAFAH1B1 \\ 
 162 & chr17 & 6920444 & 7320444 & + & ENSG00000072778 & ACADVL \\ 
 163 & chr17 & 7893564 & 8293564 & - & ENSG00000196544 & C17orf59 \\ 
 164 & chr17 & 8080029 & 8480029 & - & ENSG00000184619 & KRBA2 \\ 
 165 & chr17 & 18180051 & 18580051 & + & ENSG00000171916 & LGALS9C \\ 
 166 & chr17 & 18653658 & 19053658 & + & ENSG00000154025 & SLC5A10 \\ 
 167 & chr17 & 30269473 & 30669473 & + & ENSG00000126858 & RHOT1 \\ 
 168 & chr17 & 33500720 & 33900720 & - & ENSG00000172716 & SLFN11 \\ 
 169 & chr17 & 39259103 & 39659103 & - & ENSG00000212658 & KRTAP29-1 \\ 
 170 & chr17 & 46408359 & 46808359 & - & ENSG00000120094 & HOXB1 \\ 
 171 & chr17 & 48512138 & 48912138 & + & ENSG00000108846 & ABCC3 \\ 
 172 & chr17 & 52846088 & 53246088 & + & ENSG00000166263 & STXBP4 \\ 
 173 & chr17 & 61759295 & 62159295 & - & ENSG00000136487 & GH2 \\ 
 174 & chr17 & 72158085 & 72558085 & - & ENSG00000204347 & BTBD17 \\ 
 175 & chr17 & 77085427 & 77813550 & - & ENSG00000167281 & RBFOX3 \\ 
 176 & chr17 & 79415495 & 79815495 & - & ENSG00000182446 & NPLOC4 \\ 
 177 & chr18 & 12502776 & 12902776 & - & ENSG00000101624 & CEP76 \\ 
 178 & chr19 & 5140814 & 5540814 & - & ENSG00000105426 & PTPRS \\ 
 179 & chr19 & 8254865 & 8654865 & + & ENSG00000185236 & RAB11B \\ 
 180 & chr19 & 8278154 & 8678154 & + & ENSG00000099785 & MARCH2 \\ 
 181 & chr19 & 11346109 & 11746109 & + & ENSG00000130175 & PRKCSH \\ 
 182 & chr19 & 40524306 & 40924306 & - & ENSG00000174521 & TTC9B \\ 
 183 & chr19 & 45481495 & 45881495 & - & ENSG00000007255 & TRAPPC6A \\ 
 184 & chr19 & 49175649 & 49575649 & + & ENSG00000087074 & PPP1R15A \\ 
 185 & chr20 & 32199110 & 32599110 & + & ENSG00000101421 & CHMP4B \\ 
 186 & chr20 & 55995632 & 56395632 & - & ENSG00000124256 & ZBP1 \\ 
 187 & chr20 & 60595323 & 60995323 & - & ENSG00000101180 & HRH3 \\ 
 188 & chr20 & 60613580 & 61013580 & + & ENSG00000130703 & OSBPL2 \\ 
 189 & chr21 & 42533870 & 42933870 & + & ENSG00000183486 & MX2 \\ 
 190 & chr21 & 47152477 & 47552477 & - & ENSG00000268040 &  \\ 
 191 & chr22 & 17365844 & 17765844 & + & ENSG00000177663 & IL17RA \\ 
 192 & chr22 & 23893279 & 24293279 & - & ENSG00000187792 & ZNF70 \\ 
 193 & chr22 & 41817100 & 42217100 & - & ENSG00000100418 & DESI1 \\ 
 194 & chrX & 69279654 & 69679654 & - & ENSG00000186912 & P2RY4 \\ 
 195 & chrX & 71549366 & 71992953 & - & ENSG00000147099 & HDAC8 \\ 
 196 & chrX & 99984422 & 100384422 & - & ENSG00000182489 & XKRX \\ 
 197 & chrX & 105845910 & 106245910 & + & ENSG00000133138 & TBC1D8B \\ 
 198 & chrX & 114054540 & 114454540 & - & ENSG00000123496 & IL13RA2 \\ 
 199 & chrX & 118084542 & 118484542 & - & ENSG00000250423 & KIAA1210 \\ 
 200 & chrX & 152754465 & 153154465 & - & ENSG00000130822 & PNCK \\ 
   \hline
  \caption{List of 200 gene regions used for simulation. Positions are given with respect to GRCh37, Ensembl ID's refer to release 75 of Ensembl} 
  \label{supptable:simlist}
\end{longtable}
%}
%\afterpage{

\begin{landscape}
\begin{longtable}{cccrccc}
 \hline
  Transcription Factor & Ensembl ID & Name &  $P_{.empirical}$ & Coordinates & Disease Overlap & Band \\ 
  \hline
  \endfirsthead
  
  \multicolumn{7}{c}%
  {{\bfseries \tablename\ \thetable{} -- continued from previous page}} \\
  \hline
 Transcription Factor & Ensembl ID & Name &  $P_{.empirical}$ & Coordinates & Disease Overlap & Band \\ 
  \hline
  \endhead
  
  \hline \multicolumn{7}{r}{{Continued on next page}} \\ 
  \endfoot
  
  \hline \hline
  \endlastfoot
	IKZF3 & ENSG00000116560 & SFPQ & 7e-06 & chr1:35458749..35858749 &  & 1p34.3 \\ 
	IKZF3 & ENSG00000020129 & NCDN & 1e-06 & chr1:35823074..36223074 &  & 1p34.3 \\ 
	IKZF3 & ENSG00000126067 & PSMB2 & $>10^{-6}$ & chr1:35907445..36307445 &  & 1p34.3 \\ 
	IKZF3 & ENSG00000236887 &  & $>10^{-6}$ & chr1:113541447..113941447 & ATD, CRO, JIA, RA, SLE, T1D & 1p13.2 \\ 
	IKZF3 & ENSG00000116793 & PHTF1 & $>10^{-6}$ & chr1:114102111..114502111 & ATD, CRO, JIA, RA, SLE, T1D & 1p13.2 \\ 
	IKZF3 & ENSG00000134242 & PTPN22 & $>10^{-6}$ & chr1:114214381..114614381 & ATD, CRO, JIA, RA, SLE, T1D, VIT & 1p13.2 \\ 
	IKZF3 & ENSG00000143321 & HDGF & $>10^{-6}$ & chr1:156536717..156936717 &  & 1q23.1 \\ 
	IKZF3 & ENSG00000162889 & MAPKAPK2 & 3e-06 & chr1:206658289..207058289 & CRO, SLE, T1D, UC & 1q32.1 \\ 
	IKZF3 & ENSG00000123685 & BATF3 & 2.5e-05 & chr1:212673327..213073327 &  & 1q32.3 \\ 
	IKZF3 & ENSG00000138031 & ADCY3 & 2.5e-05 & chr2:24942708..25342708 & CRO, MS, T1D, UC & 2p23.3 \\ 
	IKZF3 & ENSG00000115137 & DNAJC27 & 1.5e-05 & chr2:24994963..25394963 & CRO, MS, T1D, UC & 2p23.3 \\ 
	IKZF3 & ENSG00000121966 & CXCR4 & 1e-06 & chr2:136675735..137075735 &  & 2q21.3 \\ 
	IKZF3 & ENSG00000163600 & ICOS & $>10^{-6}$ & chr2:204601471..205001471 & AA, ATD, CEL, PSC, RA, T1D & 2q33.2 \\ 
	IKZF3 & ENSG00000121807 & CCR2 & $>10^{-6}$ & chr3:46195225..46595225 & CEL, JIA, T1D & 3p21.31 \\ 
	IKZF3 & ENSG00000181722 & ZBTB20 & $>10^{-6}$ & chr3:114056941..115066118 &  & 3q13.31 \\ 
	IKZF3 & ENSG00000056972 & TRAF3IP2 & $>10^{-6}$ & chr6:111727481..112127481 & CRO, PSO, UC & 6q21 \\ 
	IKZF3 & ENSG00000146433 & TMEM181 & $>10^{-6}$ & chr6:158757468..159157468 &  & 6q25.3 \\ 
	IKZF3 & ENSG00000185811 & IKZF1 & $>10^{-6}$ & chr7:50143720..50543720 & CRO, MS, T1D & 7p12.2 \\ 
	IKZF3 & ENSG00000197157 & SND1 & 1e-06 & chr7:127092234..127732661 &  & 7q31.33 \\ 
	IKZF3 & ENSG00000164733 & CTSB & $>10^{-6}$ & chr8:11526957..11926957 &  & 8p23.1 \\ 
	IKZF3 & ENSG00000107249 & GLIS3 & $>10^{-6}$ & chr9:3824127..4548392 & T1D & 9p24.2 \\ 
	IKZF3 & ENSG00000165006 & UBAP1 & 2e-05 & chr9:33979003..34379003 &  & 9p13.3 \\ 
	IKZF3 & ENSG00000134453 & RBM17 & $>10^{-6}$ & chr10:5930950..6330950 & AA, ATD, CRO, JIA, MS, PSC, RA, T1D & 10p15.1 \\ 
	IKZF3 & ENSG00000170525 & PFKFB3 & $>10^{-6}$ & chr10:5986881..6386881 & AA, ATD, CRO, JIA, MS, PSC, RA, T1D & 10p15.1 \\ 
	IKZF3 & ENSG00000065675 & PRKCQ & $>10^{-6}$ & chr10:6422263..6822263 & CEL, T1D & 10p15.1 \\ 
	IKZF3 & ENSG00000183621 & ZNF438 & 8e-06 & chr10:31109136..31520866 & MS, RA & 10p11.23 \\ 
	IKZF3 & ENSG00000160584 & SIK3 & 1.9e-05 & chr11:116714118..117169153 &  & 11q23.3 \\ 
	IKZF3 & ENSG00000110344 & UBE4A & $>10^{-6}$ & chr11:118030300..118430300 & CEL, MS, PBC, RA, SJO & 11q23.3 \\ 
	IKZF3 & ENSG00000184293 & CLECL1 & $>10^{-6}$ & chr12:9685895..10085895 & MS, T1D & 12p13.31 \\ 
	IKZF3 & ENSG00000139626 & ITGB7 & 6e-06 & chr12:53401091..53801091 &  & 12q13.13 \\ 
	IKZF3 & ENSG00000185664 & PMEL & $>10^{-6}$ & chr12:56167101..56567101 & AA, PSO, T1D & 12q13.2 \\ 
	IKZF3 & ENSG00000123411 & IKZF4 & $>10^{-6}$ & chr12:56201443..56601443 & AA, PSO, T1D, VIT & 12q13.2 \\ 
	IKZF3 & ENSG00000204856 & FAM216A & $>10^{-6}$ & chr12:110706169..111106169 &  & 12q24.11 \\ 
	IKZF3 & ENSG00000111275 & ALDH2 & $>10^{-6}$ & chr12:112004691..112404691 & AS, CEL, JIA, PBC, PSC, RA, T1D & 12q24.12 \\ 
	IKZF3 & ENSG00000089022 & MAPKAPK5 & $>10^{-6}$ & chr12:112079782..112479782 & AS, CEL, JIA, PBC, PSC, RA, T1D & 12q24.12 \\ 
	IKZF3 & ENSG00000102580 & DNAJC3 & 1.1e-05 & chr13:96129393..96529393 &  & 13q32.1 \\ 
	IKZF3 & ENSG00000185650 & ZFP36L1 & $>10^{-6}$ & chr14:69063190..69463190 & CEL, CRO, JIA, MS, T1D & 14q24.1 \\ 
	IKZF3 & ENSG00000072110 & ACTN1 & 4e-06 & chr14:69246157..69646157 & CEL, CRO, JIA, MS, T1D & 14q24.1 \\ 
	IKZF3 & ENSG00000100599 & RIN3 & 2.3e-05 & chr14:92780118..93180118 &  & 14q32.12 \\ 
	IKZF3 & ENSG00000100811 & YY1 & 7e-06 & chr14:100504635..100904635 &  & 14q32.2 \\ 
	IKZF3 & ENSG00000103811 & CTSH & $>10^{-6}$ & chr15:79041916..79441916 & MS, NAR, T1D & 15q25.1 \\ 
	IKZF3 & ENSG00000179583 & CIITA & $>10^{-6}$ & chr16:10771055..11171055 & CEL, MS, PBC, PSO, T1D & 16p13.13 \\ 
	IKZF3 & ENSG00000182108 & DEXI & $>10^{-6}$ & chr16:10836317..11236317 & CEL, MS, PBC, PSO, T1D & 16p13.13 \\ 
	IKZF3 & ENSG00000168488 & ATXN2L & 6e-06 & chr16:28634356..29034356 & AS, CRO, T1D & 16p11.2 \\ 
	IKZF3 & ENSG00000184517 & ZFP1 & 2e-06 & chr16:74982390..75382390 & T1D & 16q23.1 \\ 
	IKZF3 & ENSG00000198931 & APRT & 1.2e-05 & chr16:88678352..89078352 &  & 16q24.2 \\ 
	IKZF3 & ENSG00000141012 & GALNS & 1.7e-05 & chr16:88723378..89123378 &  & 16q24.3 \\ 
	IKZF3 & ENSG00000185722 & ANKFY1 & $>10^{-6}$ & chr17:3967274..4367274 &  & 17p13.2 \\ 
	IKZF3 & ENSG00000141753 & IGFBP4 & $>10^{-6}$ & chr17:38399702..38799702 & T1D & 17q21.1 \\ 
	IKZF3 & ENSG00000108465 & CDK5RAP3 & 3e-06 & chr17:45845176..46245176 & AS, MS & 17q21.32 \\ 
	IKZF3 & ENSG00000150637 & CD226 & 1.4e-05 & chr18:67429039..67829039 & PSC, RA, T1D, UC & 18q22.2 \\ 
	IKZF3 & ENSG00000105655 & ISYNA1 & 6e-06 & chr19:18349111..18749111 & MS & 19p13.11 \\ 
	IKZF3 & ENSG00000105700 & KXD1 & $>10^{-6}$ & chr19:18468572..18868572 &  & 19p13.11 \\ 
	IKZF3 & ENSG00000105281 & SLC1A5 & 1e-06 & chr19:47091851..47491851 & PSC, T1D, UC & 19q13.32 \\ 
	IKZF3 & ENSG00000042753 & AP2S1 & $>10^{-6}$ & chr19:47154249..47554249 & PSC, T1D & 19q13.32 \\ 
	IKZF3 & ENSG00000087074 & PPP1R15A & 2e-06 & chr19:49175649..49575649 & CRO, T1D & 19q13.33 \\ 
	IKZF3 & ENSG00000104805 & NUCB1 & 1e-06 & chr19:49203307..49603307 & CRO, T1D & 19q13.33 \\ 
	IKZF3 & ENSG00000160190 & SLC37A1 & $>10^{-6}$ & chr21:43716118..44116118 & CEL, RA, T1D & 21q22.3 \\ 
	IKZF3 & ENSG00000185339 & TCN2 & $>10^{-6}$ & chr22:30802825..31202825 &  & 22q12.2 \\ 
	IKZF3 & ENSG00000128311 & TST & $>10^{-6}$ & chr22:37215681..37615681 & JIA, T1D & 22q12.3 \\ 
	IKZF3 & ENSG00000100385 & IL2RB & $>10^{-6}$ & chr22:37371094..37771094 & JIA, T1D & 22q12.3 \\ 
	IKZF3 & ENSG00000100055 & CYTH4 & $>10^{-6}$ & chr22:37478068..37878068 & JIA, T1D & 22q12.3 \\ 
	IKZF3 & ENSG00000128268 & MGAT3 & 4e-06 & chr22:39653349..40053349 & CRO, PBC, UC & 22q13.1 \\ 
	IKZF3 & ENSG00000130826 & DKC1 & $>10^{-6}$ & chrX:153791031..154191031 & T1D & Xq28 \\ 
	IKZF3 & ENSG00000130830 & MPP1 & $>10^{-6}$ & chrX:153849282..154249282 & T1D & Xq28 \\ 
	BATF & ENSG00000020633 & RUNX3 & 0.00017 & chr1:25091612..25491612 & AS, PSO & 1p36.11 \\ 
	BATF & ENSG00000236887 &  & 2e-06 & chr1:113541447..113941447 & ATD, CRO, JIA, RA, SLE, T1D & 1p13.2 \\ 
	BATF & ENSG00000162889 & MAPKAPK2 & $>10^{-6}$ & chr1:206658289..207058289 & CRO, SLE, T1D, UC & 1q32.1 \\ 
	BATF & ENSG00000203705 & TATDN3 & 0.00018 & chr1:212765170..213165170 &  & 1q32.3 \\ 
	BATF & ENSG00000152291 & TGOLN2 & $>10^{-6}$ & chr2:85355548..85755548 &  & 2p11.2 \\ 
	BATF & ENSG00000144455 & SUMF1 & 1e-06 & chr3:3742498..4708965 &  & 3p26.2 \\ 
	BATF & ENSG00000175857 & GAPT & 0.00014 & chr5:57587262..57987262 &  & 5q11.2 \\ 
	BATF & ENSG00000241685 & ARPC1A & 0.00022 & chr7:98723521..99123521 & CRO, UC & 7q22.1 \\ 
	BATF & ENSG00000197157 & SND1 & $>10^{-6}$ & chr7:127092234..127732661 &  & 7q31.33 \\ 
	BATF & ENSG00000107249 & GLIS3 & $>10^{-6}$ & chr9:3824127..4548392 & T1D & 9p24.2 \\ 
	BATF & ENSG00000184293 & CLECL1 & $>10^{-6}$ & chr12:9685895..10085895 & MS, T1D & 12p13.31 \\ 
	BATF & ENSG00000111275 & ALDH2 & $>10^{-6}$ & chr12:112004691..112404691 & AS, CEL, JIA, PBC, PSC, RA, T1D & 12q24.12 \\ 
	BATF & ENSG00000152520 & PAN3 & 0.00011 & chr13:28512643..28912643 &  & 13q12.2 \\ 
	BATF & ENSG00000185650 & ZFP36L1 & $>10^{-6}$ & chr14:69063190..69463190 & CEL, CRO, JIA, MS, T1D & 14q24.1 \\ 
	BATF & ENSG00000170291 & ELP5 & 0.00019 & chr17:6954735..7354735 &  & 17p13.1 \\ 
	BATF & ENSG00000090339 & ICAM1 & 0.00023 & chr19:10181511..10581511 & AS, CRO, JIA, MS, PBC, PSO, RA, T1D, UC & 19p13.2 \\ 
	BATF & ENSG00000160190 & SLC37A1 & $>10^{-6}$ & chr21:43716118..44116118 & CEL, RA, T1D & 21q22.3 \\ 
	BATF & ENSG00000099995 & SF3A1 & $>10^{-6}$ & chr22:30552936..30952936 & CRO, T1D, UC & 22q12.2 \\ 
	BATF & ENSG00000128311 & TST & $>10^{-6}$ & chr22:37215681..37615681 & JIA, T1D & 22q12.3 \\ 
	ESRRA & ENSG00000143321 & HDGF & $>10^{-6}$ & chr1:156536717..156936717 &  & 1q23.1 \\ 
	ESRRA & ENSG00000143479 & DYRK3 & 1e-06 & chr1:206608881..207008881 & CRO, SLE, T1D, UC & 1q32.1 \\ 
	ESRRA & ENSG00000153551 & CMTM7 & 6e-06 & chr3:32233163..32633163 &  & 3p22.3 \\ 
	ESRRA & ENSG00000186106 & ANKRD46 & 1.9e-05 & chr8:101372012..101772012 &  & 8q22.2 \\ 
	ESRRA & ENSG00000164761 & TNFRSF11B & 6.1e-05 & chr8:119764439..120164439 &  & 8q24.12 \\ 
	ESRRA & ENSG00000134453 & RBM17 & $>10^{-6}$ & chr10:5930950..6330950 & AA, ATD, CRO, JIA, MS, PSC, RA, T1D & 10p15.1 \\ 
	ESRRA & ENSG00000171206 & TRIM8 & 3.5e-05 & chr10:104204253..104604253 &  & 10q24.32 \\ 
	ESRRA & ENSG00000110651 & CD81 & $>10^{-6}$ & chr11:2197407..2597407 & T1D & 11p15.5 \\ 
	ESRRA & ENSG00000213619 & NDUFS3 & 5.1e-05 & chr11:47386888..47786888 & MS & 11p11.2 \\ 
	ESRRA & ENSG00000123444 & KBTBD4 & 8.2e-05 & chr11:47400567..47800567 & MS & 11p11.2 \\ 
	ESRRA & ENSG00000069493 & CLEC2D & 1e-06 & chr12:9617565..10017565 & MS, T1D & 12p13.31 \\ 
	ESRRA & ENSG00000110848 & CD69 & 2e-06 & chr12:9713497..10113497 & MS, T1D & 12p13.31 \\ 
	ESRRA & ENSG00000062485 & CS & 7.1e-05 & chr12:56494176..56894176 & AA, PSO, T1D & 12q13.2 \\ 
	ESRRA & ENSG00000122986 & HVCN1 & $>10^{-6}$ & chr12:110942755..111342755 &  & 12q24.11 \\ 
	ESRRA & ENSG00000089248 & ERP29 & $>10^{-6}$ & chr12:112251120..112651120 & AS, CEL, JIA, PBC, PSC, RA, T1D & 12q24.12 \\ 
	ESRRA & ENSG00000102580 & DNAJC3 & 1.5e-05 & chr13:96129393..96529393 &  & 13q32.1 \\ 
	ESRRA & ENSG00000100605 & ITPK1 & 3.6e-05 & chr14:93382665..93782665 &  & 14q32.12 \\ 
	ESRRA & ENSG00000168488 & ATXN2L & 5e-06 & chr16:28634356..29034356 & AS, CRO, T1D & 16p11.2 \\ 
	ESRRA & ENSG00000153774 & CFDP1 & 2.3e-05 & chr16:75267383..75667383 & T1D & 16q23.1 \\ 
	ESRRA & ENSG00000198931 & APRT & 5e-06 & chr16:88678352..89078352 &  & 16q24.2 \\ 
	ESRRA & ENSG00000185722 & ANKFY1 & $>10^{-6}$ & chr17:3967274..4367274 &  & 17p13.2 \\ 
	ESRRA & ENSG00000161395 & PGAP3 & 2e-06 & chr17:37653050..38053050 & CRO, MS, PBC, RA, T1D, UC & 17q12 \\ 
	ESRRA & ENSG00000141753 & IGFBP4 & $>10^{-6}$ & chr17:38399702..38799702 & T1D & 17q21.1 \\ 
	ESRRA & ENSG00000161654 & LSM12 & 7.1e-05 & chr17:41944987..42344987 &  & 17q21.31 \\ 
	ESRRA & ENSG00000167807 &  & 3.6e-05 & chr19:10226685..10626685 & AS, CRO, JIA, MS, PBC, PSO, RA, T1D, UC & 19p13.2 \\ 
	ESRRA & ENSG00000076662 & ICAM3 & 3.9e-05 & chr19:10250499..10650499 & AS, CRO, JIA, MS, PBC, PSO, RA, T1D, UC & 19p13.2 \\ 
	ESRRA & ENSG00000105281 & SLC1A5 & $>10^{-6}$ & chr19:47091851..47491851 & PSC, T1D, UC & 19q13.32 \\ 
	ESRRA & ENSG00000198053 & SIRPA & 1.3e-05 & chr20:1675154..2075154 & T1D & 20p13 \\ 
	ESRRA & ENSG00000160190 & SLC37A1 & $>10^{-6}$ & chr21:43716118..44116118 & CEL, RA, T1D & 21q22.3 \\ 
   \hline
  \caption{Genes with significant $mean(-log(p_{T1D}))$ identified from enriched gene sets. Positions are given with respect to GRCh37, Ensembl ID's refer to release 75 of Ensembl. Alopecia Areata (AA), Ankylosing Spondylitis (AS) ATD - Autoimmune thyroid disease (ATD), Celiac disease (CEL), Crohn's disease (CD), Juvenile Idiopathic Arthritis (JIA), Multiple Sclerosis (MS), Narcolepsy (NAR), Primary Biliary Cirrhosis (PBC), Primary Sclerosing Cholangitis (PSC), Psoriasis (PSO), Rheumatoid Arthritis (RA), Sjogren's syndrome (SJO), Systemic Lupus Erythematosus (SLE), Ulcerative Colitis (UC), Vitiligo (VIT)} 
  \label{supptab:siglist}         
\end{longtable} 
\end{landscape}
}

\end{document}